# Public Perceptions of Fukushima Food Products in South Korea and Its dispute Resolution : A Comparative Study on East Asia

Young Chan Seo [a*]
[a] Dept of Medical Device development, Graduate School of Medicine, Seoul National University,103 Daehak-ro, Jongno-gu, Seoul, Republic of Korea
*Corresponding author: youngchanseo@snu.ac.kr

## 1. Introduction

The differences in the perceptions of risk in Korea and Japan about Fukushima food products, which may be considered a risk of radioactive contamination, have caused legal disputes over the acceptability of food products from Fukushima prefecture. However, these disputes existed solely between Korea and Japan (World Trade Organization (WTO) 2019). There were no conflicts regarding imports of food products from Fukushima in other countries. In this paper, we speculate that the political dispute between Korea and Japan is partly attributable to the high-risk perception of Fukushima-origin food products mediated by the decision-making process among individuals.

## 2. Background and literature review: Public perception of scientific technology risk and danger

Public perception and public opinion are technically different, but we can be sure that public perception will be similar to public opinion. The impact of public opinion on policy is significant [1]. In the case of nuclear power, the public perception of scientific technology and the risks of this scientific technology may differ [2]. Therefore, the actual risks of science and technology such as nuclear power and the public's perception of risks may be different, and this gap may affect the policy-making process.

Risk is the product of an understood process and a mental process, and it is difficult to combine the two, making it difficult to integrate public awareness of risk into the policy process [3]. However, we can see the difference between the risks evaluated with objective data and the risks perceived by the public, and we can try the process of modifying or creating policies.

## 3. Survey and its assessment: Public perception in Korea about Fukushima Food Products

*3.1 Indirect and absolute assessment of public perception*

One of the reasons for the international dispute may be the distinct awareness of the Korean public, which is different from the real risk of nuclear energy.

Since the Fukushima nuclear power plant accident, the negative image of nuclear power has increased in Korea [4]. Experts can distinguish between nuclear power plant risk and radioactive food risk, while the public evaluates radiological effects such as nuclear waste, nuclear power plant accidents, and radiation on the same line when evaluating radiological risks (TanjaPerko, 2014). Therefore, we can assume that Koreans attach a negative connotation to Fukushima food products since the nuclear power plant accidents.

Compared to other energy sources, the accident rate of nuclear power plants is low [Figure. 1]. This includes the Fukushima Da-.iichi nuclear power plant accident. [Figure. 1] does not directly indicate that food products from Fukushima are not safe, but this information could affect public risk perception about Fukushima food products. This is because the public is very likely to evaluate the risks to nuclear power generation and Fukushima food on the same line.

In conclusion, since the Fukushima nuclear accident, public awareness of nuclear risks has increased, which is very likely to be considered the same risk as the Fukushima food product line to the public. Therefore, we can infer that if information about actual nuclear risks is disseminated to the public, the risk perception of Fukushima food will decrease.

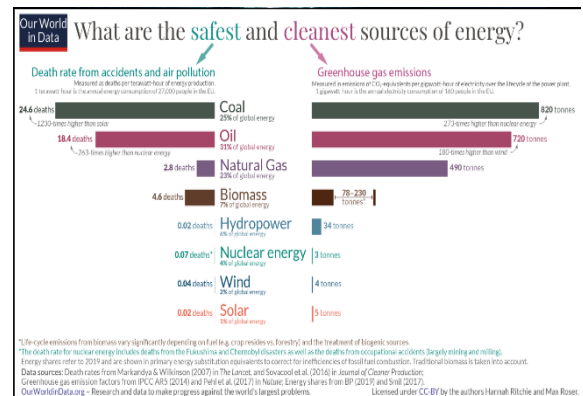

Figure. 1. < What are the safest and cleanest sources of energy?>, Our World Data, 2020 [1]

---

[1] Our world in Data, 2020, "What are the safest and cleanest sources of energy ?",

https://ourworldindata.org/safest-sources-of-energy

*3.2. Relative assessment of public perception*

Table I: Survey Status

| Survey respondents | 176 |
|---|---|
| Chinese respondents | 34 |
| Japanese respondents | 44 |
| Korean respondents | 45 |
| Others | 53 |
| Method of the survey | Google form |
| Statistic tool | Excel |
| Period | June 2020 |

Through a comparison of awareness of Korea and other countries (relative comparison), Koreans' perceptions of Fukushima and food from Japan differ greatly from those of foreign countries [Figure. 2]. Especially in the question, <4. Do you think food in Fukushima is safe from radioactive materials?>, Korea's perceptions greatly differ from other countries (China, Japan, etc.) [Figure. 2]. It means that compared to other countries such as China, Japan, and other countries (surveyed in English, Russian, and Spanish languages), more Koreans tend to perceive Fukushima food products as being dangerous due to the risk of radioactivity. The difference between China and the other countries is not significant because their P-value is more than 0.05 (significant value), so we can assume their perceptions are the same. In other cases, the P-values are different, so we can think the awareness is all different [Figure. 3]

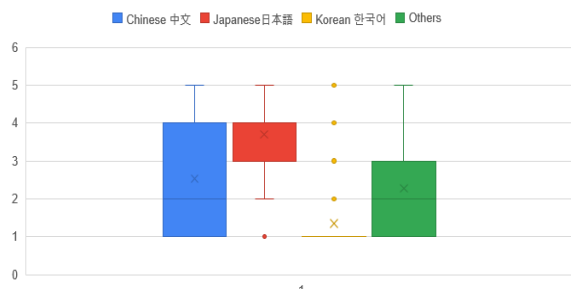

Figure. 2. Survey Questionnaire, Do you think food produced in Fukushima is safe from radioactive materials?
.

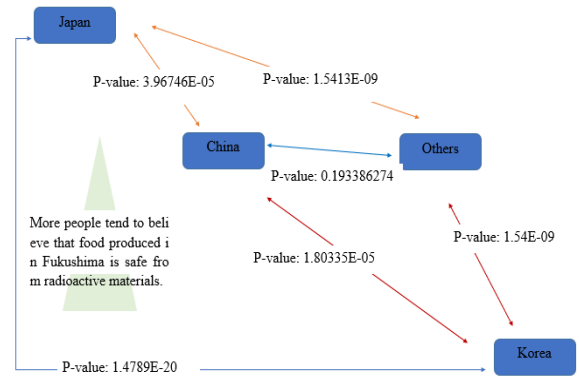

Figure. 3. Comparison plot: P-value, significance value=0.05

**4. Regression analysis**

*4.1 What variables affect the perception results?*

If the relationship between independent and dependent variables is linear, the linear regression model can represent their relationship. Common in all countries, the higher the radiation-related knowledge quiz score, the more people think Japanese food products are safe from radioactive materials, the more people think the Fukushima nuclear power plant accident handling is well, and the more people tend to think Fukushima food products are safe from radioactive materials [Table II]. Some important socio-demographic variables such as gender, age, education, and marital status can influence support or opposition to energy sources [5], but they are neglected..

Table II: Coefficients of linear regression

|  | China | Japan | Korea | Others |
|---|---|---|---|---|
| Y-Intercept | -1.340960099 | 0.800396313 | 0.650212461 | -0.775299824 |
| Quiz score | 0.174883357 | 0.051807903 | 0.111431762 | 0.086945725 |
| 1. Do you trust the information released by the Japanese government about the Fukushima nuclear accident? | 0.873814507 | -0.090109698 | 0.294801943 | -0.210178052 |
| 2. Do you trust the Japanese media's information about the Fukushima nuclear accident? | -0.300096865 | 0.74563616 | -0.068808965 | 0.273590446 |
| 3. Do you trust foreign( non-japanese) media information about the Fukushima nuclear accident? | 0.220642724 | 0.064193044 | -0.091747092 | 0.14774591 |
| 5. Do you think food produced throughout Japan is safe from radioactive materials? | 0.153318275 | 0.103793673 | 0.129634724 | 0.275972009 |
| 6. Do you think food in countries other than Japan is safe from radioactive materials? | -0.546418232 | -0.090801004 | -0.227808071 | 0.018198461 |
| 7. Do you think nuclear power is safe? | 0.166952059 | -0.021656375 | -0.051996659 | 0.118566049 |
| 8. After the Fukushima nuclear accident, do you think the treatment is going well? | 0.578229254 | 0.064300039 | 0.318849455 | 0.20094927 |

Koreans tend to believe that food products from Fukushima are more dangerous than others. While having knowledge about radiation can influence people's opinions, this does not mean that Korean people are not educated on this topic. According to the quiz score about radiation, Koreans got the highest score out of all the countries. The three groups (China, Japan, and others), did not have statistically different scores. However, this does not necessarily mean that Koreans are the most educated because the quiz questions may not have been translated well from Korean to other foreign languages, or because the survey subjects are not combined to consist of people with similar knowledge levels [Figure. 4].

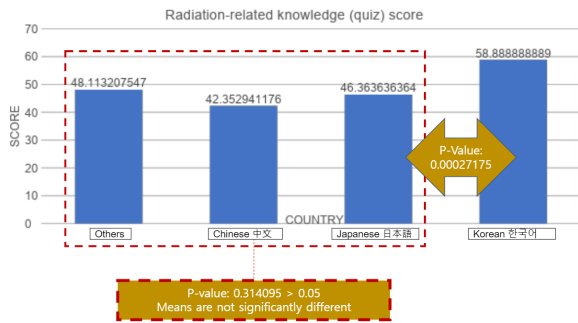

Figure. 4. Quiz Score

*4.2 Comprehensive assessment of public perception*

The indirect and absolute assessment shows that public perception of the risk of Fukushima food products could be reduced by knowing the actual risk of the nuclear power plant, and statistical survey research shows that, compared to other countries including China and Japan, Koreans tend to think agricultural products from Fukushima are dangerous. In conclusion, Koreans exceedingly think that agricultural products from Fukushima have dangerous amounts of radiation. Regression analysis [Table II] shows that with a well-handled public relations campaign, and increased public education on the effects of radiation, the negative stigma that Koreans have against food products from Fukushima could be reduced.

## 5. Reason of perception differences & international conflicts: the author's subjective opinion

Since Korea is the closest country to Fukushima, Japan, public sentiment can be affected by geographical factors. It is not clear whether the relationship between geographical distance and public perception is linear. In the case of China, it could be explained that public perception is affected by how close it is to a nuclear power plant. For example, a month after the Fukushima plant accident, the price of land near nuclear power plants in China dropped by 18 percent [6]. Furthermore, public sentiment in both Japan and Korea may have affected the international conflict over the import of Fukushima's food products. Some elements of Japanese society, such as the far-right political organization Zaitoku-kai, claim that Korea is an inferior moral country [7]. Moreover, since 1945, both Japanese and Korean leaders have created nationalist sentiments, exacerbating the divisions between the two countries [8].

## 6. Risk Perception & Decision-Making Process

Ki Yoon Sohn et al claim that Decision-Making attributes are the following [9].

Economic (cost), Safety (collective dose), Technology (degree of contribution), International affairs (diplomacy), Public risk perception (public opinion), International affairs (diplomacy) was the opinion of experts. However, the basis for this was not presented in detail.

The scholar also argued that: In the nuclear study, perceived risk-psychological dimensions are

Inequity, Not easily reducible, Risk to next-generation, Catastrophic potential, Unknown to science, The immediacy of consequences, Dread.

Therefore, in the nuclear study, it is judged that each perceived risk-psychological dimension influences the public risk perceptions and the public risk perceptions contribute to the Decision-Making Process. In the case of diplomatic or political issues, they do not affect the public perceptions but do affect the Decision-Making process [Figure. 5].).

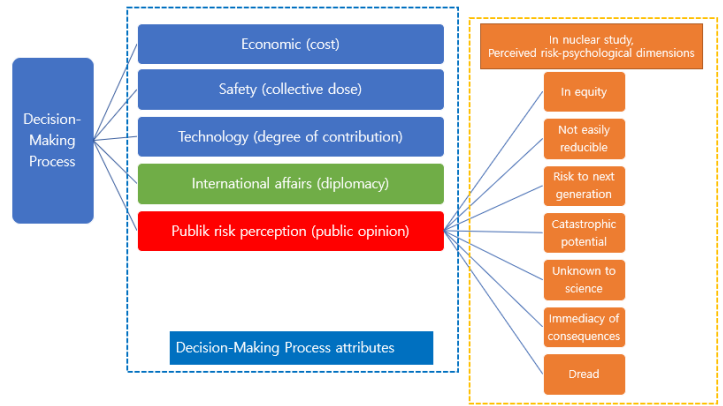

Figure. 5. Risk perception & Decision-Making Process model in nuclear study

## 7. Conclusion

With the relative perception evaluation and indirect and objective evaluation, we can infer that Korea's public risk perception of Fukushima food products is excessive. It is necessary to change the public risk perception because we want to avoid conflicts between Korea and Japan. A well-handled public relations campaign and increased public education on the effects of radiation could reduce Korean anxiety over the import of food products from Fukushima.

The conclusion does not mean to imply that the Korean people's current perception of Fukushima food products is invalid, but when compared to other countries, Korea seems to have a significantly higher negative perception. It is important to investigate further into the reasons for this, whether that be political, demographical, educational, or other factors. Another thing that should be investigated more in a follow-up study would be how people's opinions change after learning of the perceptions of other countries. It is possible one's opinion could be swayed after learning that other countries feel differently than their own.

Therefore, the government and politicians should encourage the people to access correct information, and it is necessary to consider new policies by grasping the level of risk perception of the people of other countries. In addition, politicians should try to communicate with the people to know how different their perceptions and public perceptions are.